\documentclass[preprint2]{aastex62}

\usepackage{natbib}

\usepackage{amsmath}
\usepackage{hyperref}
\usepackage{listings}
\usepackage{color}
\usepackage{float}
\usepackage{xspace} 
\usepackage{longtable}

\newcommand{\kepler}{\textsl{Kepler}\xspace}
\newcommand{\spitzer}{\textsl{Spitzer}\xspace}

\begin{document}

\title{Non-detection of Contamination by Stellar Activity in the Spitzer Transit Light Curves of TRAPPIST-1}

\author[0000-0003-2528-3409]{Brett M. Morris}
\affiliation{Astronomy Department, University of Washington, Seattle, WA 98195, USA}

\author[0000-0002-0802-9145]{Eric Agol}
\altaffiliation{Guggenheim Fellow}
\affiliation{Astronomy Department and Virtual Planetary Laboratory, University of Washington, Seattle, WA 98195, USA}

\author[0000-0003-1263-8637]{Leslie Hebb}
\affiliation{Physics Department, Hobart and William Smith Colleges, Geneva, NY 14456, USA}

\author{Suzanne L. Hawley}
\affiliation{Astronomy Department and Virtual Planetary Laboratory, University of Washington, Seattle, WA 98195, USA}

\author{Micha{\"e}l Gillon}
\author{Elsa Ducrot}
\affiliation{Space Sciences, Technologies and Astrophysics Research (STAR) Institute, Universite de Liege, Allee du 6 Aout 19C, B-4000 Liege, Belgium}

\author{Laetitia Delrez}
\affiliation{Cavendish Laboratory, JJ Thomson Avenue, Cambridge, CB3 0HE, UK}

\author{James Ingalls}
\affiliation{IPAC, California Institute of Technology, 1200 E California Boulevard, Mail Code 314-6, Pasadena, California 91125, USA}

\author{Brice-Olivier Demory}
\affiliation{University of Bern, Center for Space and Habitability, Gesellschaftsstrasse 6, CH-3012, Bern, Switzerland}

\begin{abstract}
We apply the transit light curve self-contamination technique of Morris et al.\ (2018) to search for the effect of stellar activity on the transits of the ultracool dwarf TRAPPIST-1 with 2018 \spitzer photometry. The self-contamination method fits the transit light curves of planets orbiting spotted stars, allowing the host star to be a source of contaminating positive or negative flux which influences the transit depths but not the ingress/egress durations. We find that none of the planets show statistically significant evidence for self-contamination by bright or dark regions of the stellar photosphere. However, we show that small-scale magnetic activity, analogous in size to the smallest sunspots, could still be lurking in the transit photometry undetected.
\end{abstract}

\keywords{stars: activity --- planets and satellites: fundamental parameters }

\object{TRAPPIST-1}

\section{Introduction}

TRAPPIST-1 is a system of seven approximately Earth-sized planets orbiting an M8V star \citep{Gillon2016, Gillon2017, Luger2017, Delrez2018}. It is the subject of much hope for characterization with the James Webb Space Telescope \citep{Gillon2016,Barstow2016,Morley2017,Batalha2018}, though stellar activity may complicate efforts to characterize the exoplanets \citep{Rackham2018}.

The photosphere of TRAPPIST-1 may be described as a mixture of several photospheric components with different temperatures, according to Hubble Space Telescope (HST) spectra in the analysis by \citet{Zhang2018}. \citet{Roettenbacher2017} showed that the spots evolve on the apparent rotation timescale, and comparison of \kepler and \spitzer time-dependent modulation independently suggests that there are bright (hot) spots in the photosphere \citep{Morris2018c}.  These hot spots appear to be correlated with strong flares in the K2 light curve, which calls into question the association of spot variability with stellar rotation. Recent analysis of the broadband transmission spectra of the TRAPPIST-1 planets yields a non-detection of spectral contamination by stellar activity \citep[upper limit of $200-300$ ppm in the spectra of planets b and d][]{Ducrot2018}.

In this work, we analyze the \spitzer transit light curves of TRAPPIST-1 with the ``self-contamination'' technique of \citet{Morris2018f}. The self-contamination method fits the transit light curves of planets orbiting spotted stars, allowing the host star to be a source of contaminating positive or negative flux which influences the transit depths. Accounting for the contamination potentially allows for robust inference of the exoplanet radii from the transit ingress and egress durations, even in the presence of extreme starspot distributions, like those predicted for TRAPPIST-1 by some \citep{Rackham2018, Zhang2018}. Crucially, unlike spot occultation observations \citep[see e.g.][]{Sanchis-Ojeda2011, Morris2017a}, the self-contamination technique can detect nearly-homogeneous distributions of spots throughout the transit chord of an exoplanet, or surrounding the transit chord of an exoplanet, so long as the transit chord has a different mean intensity than the rest of the photosphere \citep{Morris2018f}.

\section{Observations}

\begin{figure}
    \centering
    \includegraphics[scale=0.8]{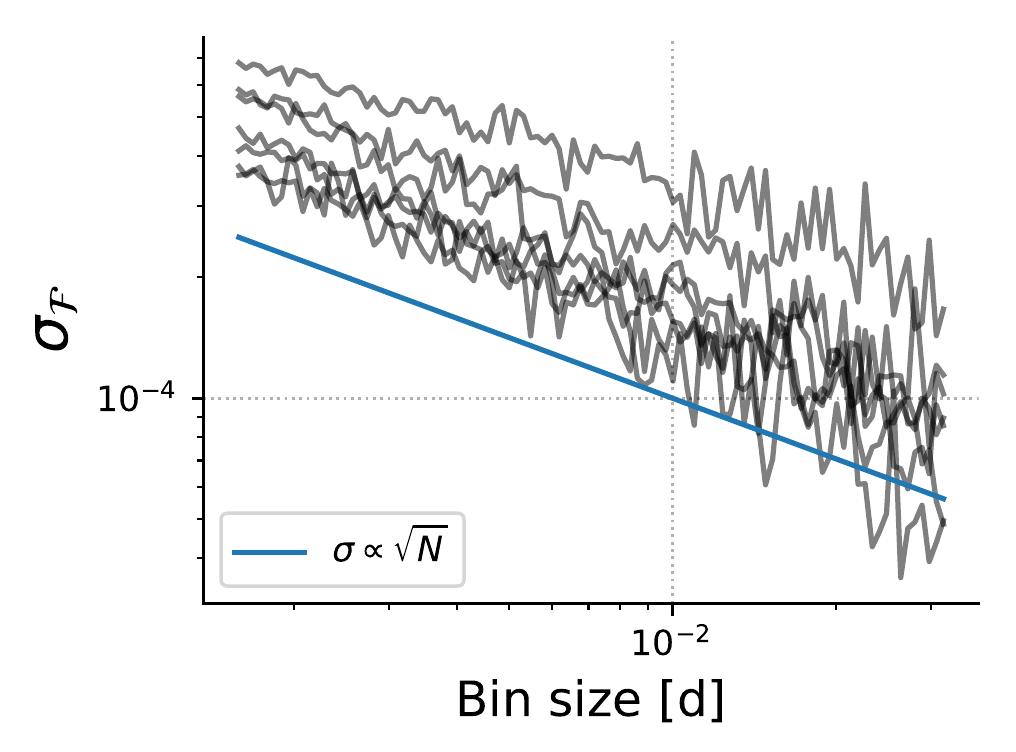}
    \caption{Lack of correlated noise in the detrended light curves of each planet (black) compared with the expectation for Gaussian, uncorrelated errors (blue).}
    \label{fig:correlatednoise}
\end{figure}

We analyze new, 2018 \spitzer observations of the TRAPPIST-1 planets including 29, 28, 16, 9, 8, 6, and 5 transits of planets b, c, d, e, f, g, and h; in \spitzer Channel 1 ($3.6 \mu$m) for planet b and Channel 2 ($4.5 \mu$m) for the others. Further detailed analysis of these data will be presented in Ducrot et al. (2018, in prep.).

We detrend each transit light curve with a linear combination of: (1) the stellar FHWM in the $\hat{x}$ and $\hat{y}$ directions; (2) the $x$ and $y$ pixel coordinates of the stellar centroid; (3) the maximum-likelihood transit model of \citet{Delrez2018} using most of their orbital parameters but allowing the mid-transit times to float; and (4) a Mat{\' e}rn-3/2 kernel Gaussian process \citep{Foreman-Mackey2017}. We vary the weights each of these observational basis vectors and fit for the mid-transit time with Markov Chain Monte Carlo (MCMC) to produce detrended light curves \citep{Foreman-Mackey2013}. The mid-transit times must be allowed to float to account for transit timing variations.

The correlated noise in the final light curve residuals is negligible after removing the Gaussian process. We verify that the noise bins down as $\propto N^{-1/2}$ by binning the noise into successively larger time bins and measuring the standard deviation of the residual flux in each bin. We find that the photometric scatter bins down to the power of $-0.55 \pm 0.08$, consistent with uncorrelated Gaussian noise, see Figure~\ref{fig:correlatednoise}. This is a weak independent confirmation that there are no significant departures from independent Gaussian uncertainties on the residual fluxes after our detrending and transit model analysis, implying a lack of occultations of bright or dark regions in the \spitzer transit light curves. In addition, we performed Anderson-Darling tests of the residuals of the \spitzer transit light curves and find that the residuals are consistent with normally distributed noise.

We then use the maximum-likelihood mid-transit times to produce phase-folded light curves of each of the planets, with the planet orbital periods from \citet{Delrez2018}. We fit the phase-folded light curve with the \citet{Morris2018f} transit light curve parameterization, which allows for significant contamination by dark starspots or regions inside or outside of the transit chord. We fit for the $p_0 = R_p/R_\star$ (the true planet radius), $p_1 \approx \sqrt{\delta}$ (where the observed transit depth is $\delta$), quadratic limb-darkening parameters, orbital inclination, mid-transit time, and semimajor axis. 

We place Gaussian priors on the semi-major axis based on the simultaneous analysis of all transits from Ducrot et al. 2018 (in prep.), which yield $\rho_\star = 50.7 \pm 2.2 \rho_\odot$ (which is related to the semi-major axis of each planet by Kepler's law), and we place Gaussian priors on the limb-darkening parameters from the global analysis which assume interpolated limb-darkening parameters from \citet{Claret2013}: $(u_1, u_2) = (0.171 \pm 0.019, 0.24 \pm 0.02)$ in \spitzer Channel 1 (3.6 $\mu $m) and $(u_1, u_2) = (0.147 \pm 0.019, 0.20 \pm 0.02)$ in Channel 2 (4.5 $\mu$m). For each planet, we also place a prior on the radius of the planet $p_0$ such that when combined with the maximum-likelihood planet mass derived by \citet{Grimm2018}, the bulk density of the planet is less than or equal to the density of iron. This last prior only informed the posterior distributions of $p_0$ for planets g and h.

\section{Results}

The results of self-contamination analysis involve comparison of the $p_0 = R_p/R_\star$ -- the true radius of the planet -- and $p_1 \approx \sqrt{\delta}$ -- the apparent transit depth \citep{Morris2018f}. Finding $p_0 < p_1$ may suggest that a relatively bright chord of the star is being occulted, while $p_0 > p_1$ may suggest that a dark chord is being occulted, or that the planet is oblate. 

\begin{figure}[H]
    \centering
    \includegraphics[scale=0.83]{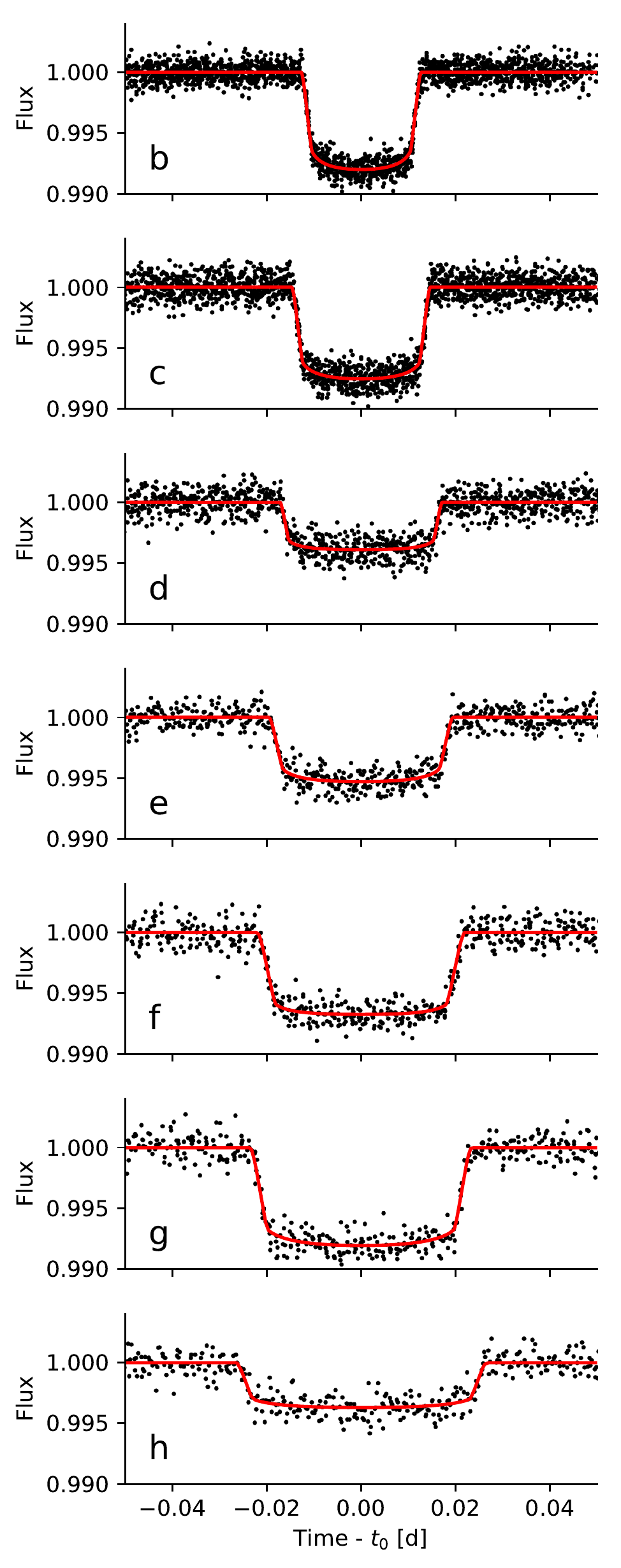}
    \caption{Maximum-likelihood transit models and detrended \spitzer light curves for each of the TRAPPIST-1 planets.}
    \label{fig:transits}
\end{figure}

\begin{figure}[H]
    \centering
    \includegraphics[scale=0.8]{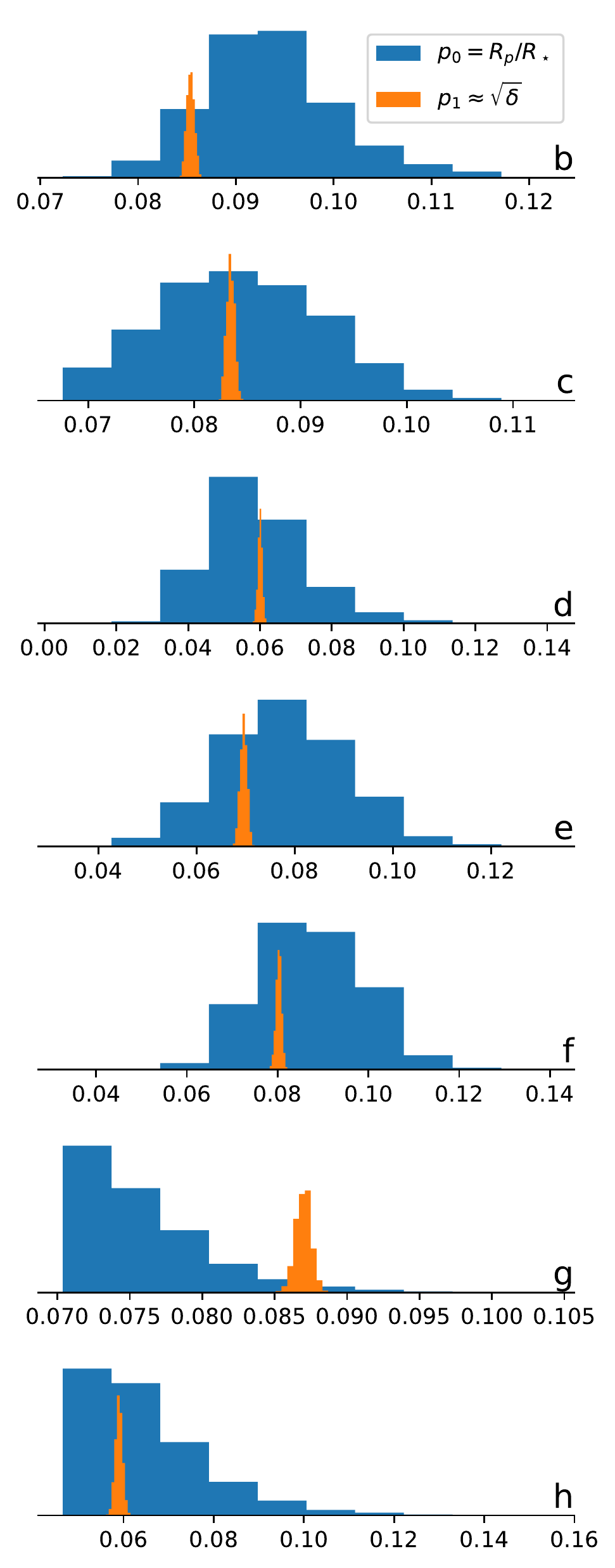}
    \caption{Posterior distributions for $p_0$ and $p_1$ for planets b through h. All of the planets present $p_0 \approx p_1$, suggesting that the photospheric intensity within transit chords of the planets is roughly representative of the rest of the stellar photosphere.}
    \label{fig:posteriors}
\end{figure}

In Figures~\ref{fig:transits} and \ref{fig:posteriors} we show the maximum-likelihood transit models, and the posterior distributions for $p_0$ and $p_1$ for each planet. For all planets, we find insignificant evidence for $p_0 \neq p_1$. The most significant discrepancy is for planet g, which is still 2-$\sigma$ consistent with the null hypothesis ($p_0 = p_1$). For planet b, the planet with the most-sampled light curve, we measure self-contamination $\epsilon = 1 - (p_1/p_0)^2 = 0.22 \pm 0.10$. 

Planet g has $p_0 < p_1$ at 94\% confidence for a contamination parameter $\epsilon = 1 - (p_1/p_0)^2 = -0.35 \pm 0.15$, though the improvement in the fit compared to one where $p_0 = p_1$ is only $\Delta \chi \approx 5$, providing insignificant evidence for occultation of TRAPPIST-1 g by a bright latitude. If more data confirm that $p_0 < p_1$, the radius of the planet \citep[from, for example,][]{Delrez2018} may be somewhat overestimated in the literature. The posterior distribution for $p_0$ has a mode near $p_0=R_p/R_\star = 0.07$, which corresponds to the limit where the planet would require the bulk density of iron. We emphasize that this result is statistically insignificant, and more observations by \spitzer or with JWST are needed to confirm or falsify the apparently small radius of planet g.

At present the observational uncertainties are too large to place robust constraints on whether or not significant spot distributions contaminate the \spitzer transit light curves using the self-contamination technique. This result is in agreement with the recent analysis by \citet{Ducrot2018}, who find insufficient evidence for strong contamination of the broad-band spectrum of the TRAPPIST-1 planets by stellar activity. The lack of apparent spot contamination is at odds with the large spot coverage upper-limits inferred by \citet{Rackham2018,Zhang2018}. Short-cadence JWST/NIRSpec observations of the TRAPPIST-1 planets will be ideal for placing more informative constraints on possible self-contamination by stellar activity on TRAPPIST-1 \citep{Batalha2018}.

Perhaps an even stronger non-detection of spots than the lack of self-contamination is the lack of active region occultations, implying an apparently facula/spot-free surface of TRAPPIST-1 within the transit chords. This spot-free region in the photosphere may seem to be at odds with the literature, which suggests that most late type M dwarfs are chromospherically active, e.g., when observed via H$\alpha$ emission \citep{West2008, West2015}. However, H$\alpha$ observations of TRAPPIST-1 in particular show $\log L_{H\alpha}/L_\mathrm{bol} = -4.70$ \citep{Reiners2018}, making this star appear relatively inactive even among late M dwarfs (see for example Figure 7 of \citealt{West2015}). We note that chromospheric activity, which influences the FUV and Ly-$\alpha$ environments of these potentially habitable worlds, is not constrained by these observations -- we are only probing the near-IR spot coverage with the self-contamination technique.

If we assume that we are viewing TRAPPIST-1 equator-on and that the orbits are aligned with the stellar spin axis ($i_\star=90^\circ$ and $\lambda = 0^\circ$), then the planets should occult latitudes from the equator up to $30^\circ$ in  one hemisphere -- which is perhaps a surprisingly small portion of the stellar hemisphere, see Table~\ref{tab} for the range of latitudes occulted by each planet. Active latitudes on young M dwarfs have been observed via Doppler imaging of HK Aqr and RE 1816 +541 \citep{Barnes2001}. HK Aqr's spot distribution peaks near $30^\circ$ latitude and most spots on RE 1816 +541 were found at latitudes $>30^\circ$ -- making most of those observed spots just out-of-reach of the transit chords of the TRAPPIST-1 planets. However, these stars are much younger and hotter than the inferred age and temperature for TRAPPIST-1 \citep{Burgasser2015}, and may not be fully convective, making it unclear if it is sensible to compare them with TRAPPIST-1.

Perhaps one way to place spots on the star without affecting the transit light curves is to place the spots at or near the poles -- it is also possible that there is a spot at one rotational pole which is hidden from view due to stellar inclination. We expect polar magnetic active regions on late M stars as they have been observed widely with Doppler imaging \citep[see e.g.][]{Strassmeier2002, Morin2008, Morin2010}. Such polar spots may be long lived; for example, the observed polar spot on the fully convective dwarf V374 Peg is stable on one-year timescales \citep{Morin2008}.  However, these will lead to rotational variability, which is constrained by the K2 dataset.  If the correlation between flares and spot brightening in the K2 data is confirmed \citep{Morris2018c}, then this may suggests that spots are not responsible for the quasi-periodic brightening.

\begin{figure*}
    \centering
    \includegraphics[scale=0.8]{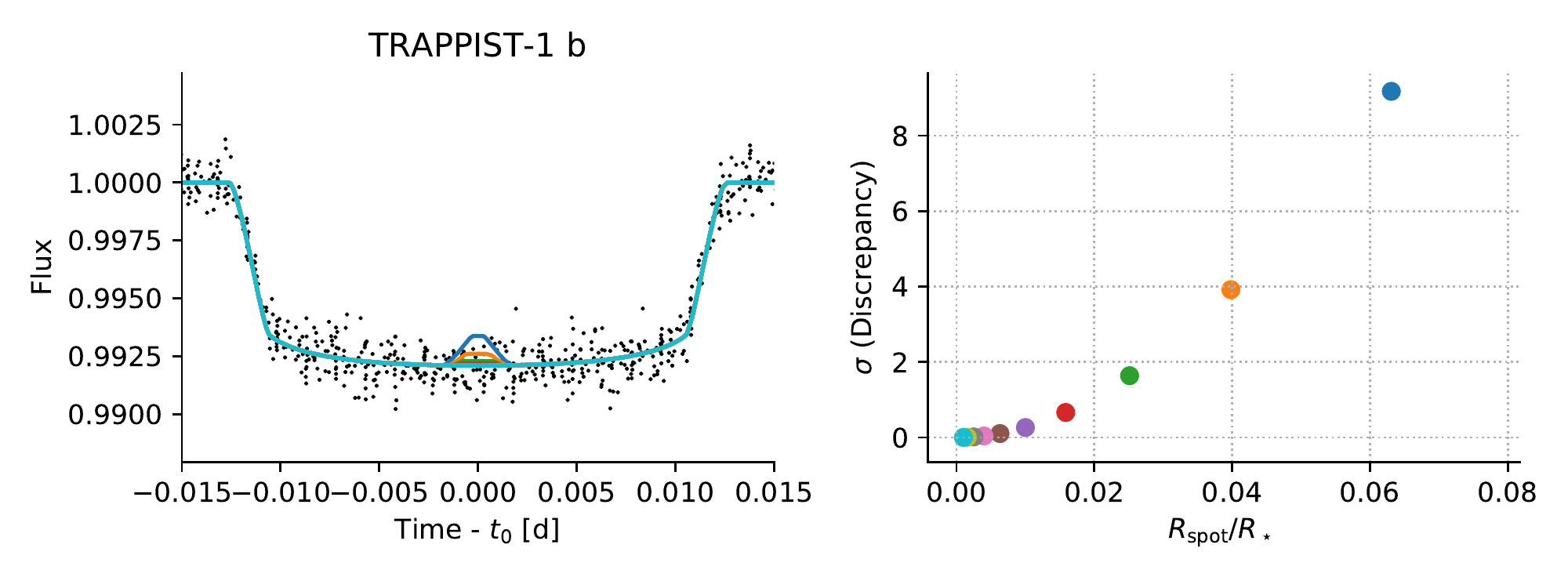}
    \caption{{\it Left:} \texttt{STSP} model light curves occulting spots of various radii and the \spitzer photometry of TRAPPIST-1 b. {\it Right:} The corresponding discrepancy between the data and the \texttt{STSP} model for each radius. We find that the photometry is insensitive to spots smaller than $R_\mathrm{spot}/R_\star < 0.04$, which is similar in size to the smallest sunspots -- so small-scale magnetic activity may still be lurking undetected within the transit chords.}
    \label{fig:stsp}
\end{figure*}

Finally, we confirm the non-detection of spot occultations in the transit light curve by modelling the light curve of TRAPPIST-1 b with \texttt{STSP} (Hebb et al.~2018\footnote{Open source, available online: \url{https://github.com/lesliehebb/stsp}}), assuming the temperature contrast of \citet{Rackham2018} ($T_\mathrm{phot} = 2500$ K, $T_\mathrm{spot} = 2064$ K; Spitzer IRAC-1 contrast $c=0.75$). We vary the radius of the spot to place an upper-limit on the plausible size of occulted spots that may go undetected in the transit photometry -- see Figure~\ref{fig:stsp}. We find that spots with radii $R_\mathrm{spot}/R_\star < 0.04$ induce spot occultations with amplitudes $\lesssim 3\sigma$ discrepant with the observed \spitzer light curve of TRAPPIST-1 b. This implies that spots within the transit chord should have physical radii of $\lesssim 3.3$ Mm. For comparison with the Sun, the smallest sunspots are roughly 1.75 Mm in radius \citep{Solanki2003}. Thus it is still possible that very small-scale magnetic activity is occurring within the transit chords, to which we are still insensitive with \spitzer's  photometric precision.

\begin{deluxetable}{lcc}
\tablecaption{Stellar latitude ranges occulted by each planet, assuming the stellar inclination is $i_\star = 90^\circ$ and the projected spin-orbit angle is $\lambda = 0^\circ$, given the impact parameters and planetary radii from \citet{Delrez2018}. \label{tab}}
\tablehead{\colhead{Planet} & \colhead{Lower [$^\circ$]} & \colhead{Upper [$^\circ$]}}
\startdata
b & $4 \pm 4$ & $14 \pm 4$ \\
c & $4 \pm 5$ & $13 \pm 5$ \\
d & $1 \pm 6$ & $8 \pm 6$ \\
e & $10 \pm 3$ & $18 \pm 3$ \\
f & $15 \pm 2$ & $25 \pm 3$ \\
g & $19 \pm 2$ & $30 \pm 2$ \\
h & $19 \pm 2$ & $27 \pm 3$
\enddata
\end{deluxetable}

\section{Conclusions}

We present a self-contamination analysis of the \spitzer transit light curves of the TRAPPIST-1 planets, using the transit light curve parameterization of \citet{Morris2018f}. We find insufficient evidence for contamination by bright or dark spots inside or outside of the transit chord using the self-contamination technique of \citet{Morris2018f}, measuring contamination $\epsilon = 0.22 \pm 0.10$ for planet b. This is a tighter constraint on the contamination than measured with \kepler photometry in \citet{Morris2018f}. This analysis suggests that the mean photosphere is similar to the photosphere occulted by the TRAPPIST-1 planets. However, we cannot exclude the possibility that small-scale magnetic activity analogous in size to the smallest sunspots may be occuring within (or outside) of the transit chords given the photometric precision of the \spitzer observations.   

\facilities{Spitzer}

\software{\texttt{astropy} \citep{Astropy2018}, \texttt{emcee}, \citep{Foreman-Mackey2013}, \texttt{ipython} \citep{ipython}, \texttt{numpy} \citep{VanDerWalt2011}, \texttt{scipy} \citep{scipy},  \texttt{matplotlib} \citep{matplotlib}, \texttt{robin} \citep{Morris2018f}}

\acknowledgements

We are grateful for the greater Spitzer TRAPPIST analysis team, without whom this analysis would not have been possible: Artem Burdanov, Valerie Van Grootel, Didier Queloz, Amaury Triaud, Julien de Wit, Adam Burgasser, Sean Carey, Sue Lederer, Emeline Bolmont, Jeremy Leconte, Sean Raymond, and Franck Selsis.
LD acknowledges support from the Simons Foundation (PI Queloz, grant number 327127) and the Gruber Foundation Fellowship. MG is a F.R.S.-FNRS Senior Research Associate. The research leading to these results has received funding from the European Research Council under the FP/2007-2013 ERC Grant Agreement n° 336480, from the ARC grant for Concerted Research Actions, financed by the Wallonia-Brussels Federation, and from the Balzan Foundation.  E.A. acknowledges NSF grant AST-1615315, NASA grant NNX14AK26G and from 
the NASA Astrobiology Institute's Virtual Planetary Laboratory Lead Team, 
funded through the NASA Astrobiology Institute under solicitation NNH12ZDA002C 
and Cooperative Agreement Number NNA13AA93A.

\bibliography{bibliography}

\end{document}